\documentclass[aps,eqsecnum,nofootinbib,superscriptaddress,showpacs]{revtex4}
\usepackage[dvips]{graphicx}  
\usepackage{amsmath,amsfonts}
\usepackage{bm}
\usepackage{color}  
\newcommand{\half}{ \frac{1}{2} }

\newcommand{\hx}{ \Hat{x} }

\newcommand{\ket}[1]{\left|~#1~\right\rangle}

\begin{document}
\title{Heterotic String in a Constant Magnetic Field}
\author{Akira Kokado}
\email{kokado@kobe-kiu.ac.jp}
\affiliation{Kobe International University, Kobe 658-0032, Japan}
\author{Gaku Konisi}
\email{konisigaku@nifty.com}
\affiliation{Department of Physics, Kwansei Gakuin University,
Sanda 669-1337, Japan}
\author{Takesi Saito}
\email{tsaito@k7.dion.ne.jp}
\affiliation{Department of Physics, Kwansei Gakuin University,
Sanda 669-1337, Japan}
\date{\today}
\begin{abstract}
When a charged heterotic string is placed in a constant magnetic field $B$,
 we show that this system can be solved exactly by using the cyclotron 
frequency. We then calculate anomalies of the super Virasoro algebra, and 
give the corresponding spectrum-generating algebra for this system. 
They differ from the free case by the cyclotron frequency. It is remarkable 
that our system is equivalent to the completely free system when $B$ takes integral values. 
\end{abstract}
\maketitle
\section{Introduction}\label{sec:intro}
The heterotic string\cite{ref:Gross_MR} is regarded as one of the most promising model for realistic particles. 
It is a closed string, which is composed of right-moving fermionic and bosonic strings with ten dimensions 
and left-moving bosonic string with 26 dimensions. As a result of the compactification of the 
26-10=16-dimensional internal space, the heterotic string is associated with the phenomenologically 
promising gauge group $SO(32)$ or $E_8\times E_8$\cite{ref:Gross_MRY,ref:Candelas_HSWMS}. \\ 
\indent The electromagnetic interaction of the heterotic string has been also considered. The electromagnetic 
field has so far been introduced as a vector field $g_{\mu I}$ of the Kaluza-Klein type, where $g_{\mu I}$ is a space-time 
metric with external components $\mu =0, \cdots, 9$ and internal components $I =10, \cdots, 25$\cite{ref:Green_SWKSST}. However, this type of electromagnetic 
interaction cannot be solved exactly, but we can treat it only by a perturbative method.  \\
\indent In this paper we propose an exactly solvable model of the heterotic string placed in a constant magnetic field. 
The electric charge was assumed to distribute uniformly along the closed string. The electromagnetic field is 
introduced in such a way that the interaction is invariant under the superconformal transformation and also 
under the gauge transformation of the electromagnetic field. When the charged heterotic string is placed 
in the constant magnetic field $B$, we show that the Lagrangian of the interacting heterotic string can be 
translated into the free type of Lagrangian. However, this free type of closed string is not periodic 
at the boundary $\sigma =2\pi $, but yields the phase factors $\exp(\pm 2\pi i\omega )$, where $\omega $ is the cyclotron frequency. This causes 
the fact that the cyclotron frequency $\omega $ is included in all orders in mode expansions of the string 
and also in quantization conditions for mode operators. This differs from the completely free case. 
Therefore, our next task is to calculate the superconformal algebra together with anomalies, 
and also to give the spectrum-generating algebra\cite{ref:Brower_F}, which is necessary to construct physical 
states satisfying the Virasoro conditions. Finally we point out that our system is equivalent 
to the completely free system when $B$ takes integral values. \\
\indent In Sec.\ref{sec:2} we propose a new type of interaction of the electromagnetic field with 
the heterotic string. In Sec.\ref{sec:3} and Sec.\ref{sec:4} we calculate anomalies associated with 
the super Virasoro algebra. In Sec.\ref{sec:5} the spectrum-generating algebra is constructed. 
In Sec.\ref{sec:6} we consider the algebra isomorphic to the spectrum-generating algebra. 
From the equation of isomorphisms we derive the number of space-time dimensions 
together with constraint constants of the 0-mode Virasoro operators. Finally, Sec.\ref{sec:conclude} is devoted to concluding remarks.
%
\section{A heterotic string in the external field}\label{sec:2}
The free Lagrangian is given by
\begin{align}
   L_0 = 2\partial _+ \hx_a D\hx^a~.
  \label{eq:Lag1}
\end{align}
The notation is summarized as follows:
\begin{align}
 & (\tau , \sigma )=(s^0, s^1), \quad s^{\pm}=s^0 \pm s^1, \quad 
\partial _{\pm} = \partial /\partial s^{\pm} = 
(1/2)(\partial /\partial \tau \pm \partial /\partial\sigma ), 
\label{eq:notion1} \\
 & D=i\partial _\theta + \theta \partial _{-}
 \nonumber
\end{align}
and
\begin{align}
 &  \hx^a(s^+, s^-, \theta ) = x^a(s^+, s^-) + i\chi ^a(s^+, s^-)\theta ~,
 \nonumber \\
 &  a = (\mu ,I), \quad \mu = 0,1, \cdots, 9 \quad I=10, \cdots, 25
   \label{eq:notion2}
\end{align}
The free action is
\begin{align}
   S_0 = 2 \int d^2 s d\theta ~\partial _{+} \hx_a D\hx^a 
  = 2 \int d^2 s~(\partial _+ x_a \partial _- x^a + i\chi _a \partial _+ \chi ^a)~.
  \label{eq:action1}
\end{align}
Under the superconformal transformation (SCT)
\begin{align}
 &  \delta s^+ = f^+(s^+)~,
 \label{eq:SCT1} \\
 &  \delta s^- = f^-(s^-) + i\eta (s^-)\theta ~, 
 \nonumber \\
 &  \delta \theta = \eta (s^-) + \half \theta \partial _{-}f^{-}(s^{-})~,
 \nonumber
\end{align}
the action is invariant, if $\hx^a$ is a superconformal scalar. The conserved superconformal charge is given by  
\begin{align}
   Q = \int d\sigma (f^+ L_+ + f^-L_- + i\eta G_-)~,
 \label{eq:SCT2}
\end{align}
where
\begin{align}
 &  L_+ = J_a J^a~, \quad J^a = \partial _+ x^a~,
 \label{eq:defL+} \\
 &  L_- = I_\mu I^\mu  + i\partial _\sigma \chi _\mu \chi ^\mu ~, \quad I^\mu = \partial _-x^\mu ~,
 \label{eq:defL;} \\
 &  G_- = 2\chi _\mu I^\mu ~.
 \label{eq:defG-}
\end{align}

\indent Now, the electromagnetic field $A_a(\hx)$ is introduced in such a way as
\begin{align}
   \Hat{L} = 2\big[\partial _{+} \hx_a + 2 A_{a}(\hx )\big]D\hx ^{a}~.
 \label{eq:LagragianEM1}
\end{align}
If the action for the interaction term is integrated over $\theta $, we have
\begin{align}
 &  S_{int} = 4 \int d^2s d\theta~A_{a}(\hx )D\hx ^{a} 
 \nonumber \\
 &          = 4 \int d^2s~\big[A_{a}(x) \partial _{-} x^{a} 
- i\partial _{a}A_{b}(x)\chi ^{a} \chi ^{b} \big]~.
 \label{eq:action2}
\end{align}
The interaction is clear to be invariant under the gauge transformation,
 $\delta A_a(x)=\partial _a \Lambda (x)$, or equivalently,
\begin{align}
   \delta A_a(\hx )=\partial _a \Lambda (\hx )
 \label{eq:gaugeA}
\end{align}
We then choose the symmetric gauge
\begin{align}
   A_{\mu }(\hx ) = - \half F_{\mu \nu } \hx^{\nu }~, \quad F_{\mu \nu } = \mbox{constant}~,
 \label{eq:sysgauge}
\end{align}
to obtain
\begin{align}
  \Hat{L} = 2\big[\partial \hx - F \cdot \hx \big]\cdot D \hx ~.
 \label{eq:L2}
\end{align}
\indent If we use a new variable
\begin{align}
  \Hat{X} = \exp(-Fs^{+})\hx ~,
 \label{eq:newvariable}
\end{align}
we find that the Lagrangian (\ref{eq:L2}) reduces to
\begin{align}
   \Hat{L} = 2 \partial _{+} \Hat{X}\cdot D \Hat{X}~.
  \label{eq:Lagrangian2}
\end{align}
This is the free type of Lagrangian with respect to $\Hat{X}$. The action for Eq.(\ref{eq:Lagrangian2}) is invariant under SCT, if   is a superconformal scalar. In this case the original variable $\hx$ behaves as
\begin{align}
  \delta _{SCT} \hx = \delta s^{+} F\cdot \hx
 \label{eq:SCT3}
\end{align}
under SCT.  \\
\indent We concentrate on one $2\times 2$ block of $F_{\mu \nu }$ with $B$ real
\begin{align}
  F_{\mu \nu }
  = \begin{pmatrix}
              0 &  B \\
             -B & 0
    \end{pmatrix}~, \quad \mu, \nu = 1, 2~.
 \label{eq:defF}
\end{align}
Introducing
\begin{align}
   \Hat{X}^{(\pm)} = \big( \Hat{X}^{1} \pm i\Hat{X}^{2}\big)/\sqrt{2}~.
 \label{eq:defX_pm}
\end{align}
Eqs.(\ref{eq:newvariable}) turns out to be
\begin{align}
  \Hat{X}^{(\pm)} = \exp{(\pm iBs^{+})}\hx^{(\pm)}~.
 \label{eq:X_pm2}
\end{align}
Corresponding to Eq.(\ref{eq:notion2}), we expand $\Hat{X}^{(\pm)}(\tau ,\sigma ,\theta )$ in $\theta $
\begin{align}
  \Hat{X}^{(\pm)}(\tau ,\sigma ,\theta ) = X^{(\pm)}(\tau ,\sigma ) + i\psi ^{(\pm )}(\tau ,\sigma )\theta~.
 \label{eq:expandX_pm1}
\end{align}
From Eq.(\ref{eq:X_pm2}) we find
\begin{align}
  X^{(\pm)}(\tau ,\sigma ) = \exp{\big[ \pm iB(\tau + \sigma )\big]}x^{(\pm)}(\tau , \sigma )~, 
 \label{eq:Xpm3} \\
  \psi ^{(\pm )}(\tau ,\sigma ) = \exp{\big[ \pm iB(\tau + \sigma )\big]}\chi ^{(\pm)}(\tau , \sigma )~.
 \label{eq:Psi3}
\end{align}
Considering the periodicity of $x^{(\pm)}(\tau , \sigma )$ and $\chi ^{(\pm)}(\tau , \sigma )$, 
we find the quasi-periodicity for $X^{(\pm)}(\tau ,\sigma )$ and $\psi ^{(\pm )}(\tau ,\sigma )$ as
\begin{align}
 & X^{(\pm)}(\tau ,\sigma + 2\pi ) = \exp{\big( \pm 2\pi iB\big)}X^{(\pm)}(\tau , \sigma )= \exp{\big( \pm 2\pi i\omega \big)}X^{(\pm)}(\tau , \sigma )~, 
 \label{eq:Xperiodicity} \\
 & \psi ^{(\pm)}(\tau ,\sigma + 2\pi ) = -\exp{\big( \pm 2\pi i\omega \big)}\psi ^{(\pm)}(\tau , \sigma )~, \quad \mbox{for NS sector}
 \label{eq:PsiperiodicityNS} \\
 & \psi ^{(\pm)}(\tau ,\sigma + 2\pi ) = \exp{\big( \pm 2\pi i\omega \big)}\psi ^{(\pm)}(\tau , \sigma )~, \quad \mbox{for Ramond sector}
 \label{eq:PsiperiodicityR}
\end{align}
where $B=q+\omega $, $q\in Z$, $0\leq \omega < 1$. In the following we call $\omega $ simply the cyclotron frequency, 
and consider only the range $0<\omega <1$, since our system is equivalent to the free system when $\omega =0$. 
However, it is remarkable that when $B$ takes integral values, $B=q\in Z$, our system is equivalent to the completely free system. \\

The relevant parts in Lagarangian (\ref{eq:Lagrangian2}) can be written as
\begin{align}
   \Hat{L} = 2\big[ \partial _{+} \Hat{X}^{(+)} D\Hat{X}^{(-)} + \partial _{+} \Hat{X}^{(-)} D \Hat{X}^{(+)} \big ]~.
 \label{eq:redefL}
\end{align}
Integrating over $\theta $, the Lagrangian reduces to
\begin{align}
  L=2\big[ \partial _{+}X^{(+)} \partial _{-}X^{(-)} + \partial _{+}X^{(-)} \partial _{-}X^{(+)} + i\psi ^{(+)}\partial _{+}\psi ^{(-)} + i\psi ^{(-)}\partial _{+}\psi ^{(+)} \big ]~.
 \label{eq:L3}
\end{align}
In spite of the quasi-periodicity of $X$ and $\psi $, the periodic boundary condition, which is necessary 
in the variational principle, is guaranteed, because the aperiodic phase factors $\exp{(\pm 2\pi i\omega )}$ 
are always cancelled out between the $(+)$ and $(-)$ components in the Lagrangian. \\
   Equations of motion are all of free type:
\begin{align}
   \partial _{+} \partial_{-} X^{(\pm)} = 0~,
 \label{eq:eq_of_motionX} \\
   \partial_{+} \psi ^{(\pm)} = 0~.
 \label{eq:eq_of_motionPsi}
\end{align}
Their solutions with boundary conditions (\ref{eq:Xperiodicity}), (\ref{eq:PsiperiodicityNS}) and (\ref{eq:PsiperiodicityR}) are given by
\begin{align}
 & X^{(\pm)}(\tau ,\sigma ) = {X_{+}}^{(\pm)}(s^{+}) + {X_{-}}^{(\pm)}(s^{-}) ~,
  \label{eq:expandX2} \\
 & {X_{+}}^{(\pm )}(s^{+}) = i \sum _{n}\frac{1}{n\mp \omega } \exp{[-i(n\mp \omega )s^{+}]} {\alpha _n}^{(\pm)}~,
  \nonumber \\
 & {X_{-}}^{(\pm)}(s^{+}) = i \sum _{n}\frac{1}{n\pm \omega } \exp{[-i(n\pm \omega )s^{+}]} {\beta _n}^{(\pm)}~,
  \nonumber
\end{align}
and
\begin{align}
 & \psi ^{(\pm)}(s^{-}) = \half \sum _{r\in Z+\half}{b_r}^{(\pm)}\exp{[-i(r\pm \omega )s^{-}]}~, \quad \mbox{for NS sector}
  \label{eq:expandX3} \\
 & \psi ^{(\pm)}(s^{-}) = \half \sum _{n\in Z}{d_n}^{(\pm)}\exp{[-i(n\pm \omega )s^{-}]}~, \quad \mbox{for Ramond sector}~.
  \nonumber
\end{align}
\indent  The conjugate momenta to $X^{(\pm)}(\tau , \sigma )$ are
\begin{align}
 &  P^{(\mp)}= \frac{\partial L}{\partial \big(\partial _\tau X^{(\pm)}\big )} = \partial _{-} X^{(\mp)} + \partial _+ X^{(\mp)} = \dot {X}^{(\mp)}~.
 \label{eq:defP}
\end{align}
The quantization is accomplished by setting the commutation rules
\begin{align}
  & \big[\, X^{(\pm)}(\tau , \sigma )\,,~P^{(\mp)}(\tau , \sigma ')\, \big] = i\, \pi \delta _{\pm\omega }(\sigma -\sigma ')~,
\label{eq:def-XP-CCR}
\end{align}
and other combinations are zero. Here $\delta _{\pm\omega }(\sigma -\sigma ')$ is the delta function with the same quasi-periodicity 
as $X^{(\pm)}(\tau , \sigma )$  with respect to its argument. From these we have commutation relations:
\begin{align}
 & \big[\, {\alpha }_m^{(+)}\,,~{\alpha }_n^{(-)}\, \big] = (m-\omega )\delta _{m+n,0}~, \quad 0<\omega <1
\label{eq:def_alpha=ccR} \\
 & \big[\, {\beta }_m^{(+)}\,,~{\beta }_n^{(-)}\, \big] = (m+\omega )\delta _{m+n,0}~,
\label{eq:def-beta-CCR}
\end{align}
and other combinations are zero. Since $0<\omega <1$, $\alpha _m^{(+)}$, $\beta _m^{(-)}$  are annihilation operators 
for $m>0$, and creation operators for $m\leq 0$, while $\alpha _m^{(-)}$, $\beta _m^{(+)}$  are creation operators for $m<0$, 
and annihilation operators for $m\geq 0$. As for the fermionic parts, we get, after the Dirac quantization,
\begin{align}
 & \big\{\, b_r^{(+)}\,,~b_s^{(-)}\, \big\} = \delta _{r+s,0}~, \quad others =0~,
\label{eq:def_b=ccR} \\
 & \big\{\, d_m^{(+)}\,,~d_n^{(-)}\, \big\} = \delta _{m+n,0}~, \quad others =0~.
\label{eq:def-d-CCR}
\end{align}
${b_r}^{(\pm)}$ are annihilation operators for $r>0$, and creation operators for $r<0$. \\ 
\indent For the Ramond sector, we need a special care on the 0-modes, so the detail will be discussed in Sec.4.
\section{Calculation of anomaly} \label{sec:3}
The Lagrangian (\ref{eq:redefL}) happens to appear as if it is a free Lagrangian. However, 
the dynamical variables $X^{(\pm)}(\tau , \sigma )$ and $\psi ^{(\pm)}(\tau , \sigma )$ are subject to 
the quasi-periodicity (\ref{eq:Xperiodicity})-(\ref{eq:PsiperiodicityR}), and this causes the inclusion of the cyclotron frequency 
$\omega $ in the commutators (\ref{eq:def_alpha=ccR}) and (\ref{eq:def-beta-CCR}) for mode operators, different from the completely free case. 
Considering this fact, we should examine the validity of the super Virasoro algebra together with calculation of anomaly. \\
\indent Let us define the current operators by
\begin{align}
 & J^{(\pm)}(z) = i\partial X_{+}^{(\pm)} = z^{\pm\omega } \sum_n z^{-n-1} \alpha _n^{(\pm)} = z^{\pm\omega }\tilde {J}^{(\pm)}(z)~, \quad z=\exp{(is^{+})}~,
\label{eq:defJ1} \\
 & I^{(\pm)}(z) = i\partial X_{-}^{(\pm)} = z^{\mp\omega } \sum_n z^{-n-1} \beta _n^{(\pm)} = z^{\mp\omega }\tilde {I}^{(\pm)}(z)~, \quad z=\exp{(is^{-})}~.
\label{eq:defI1}
\end{align}  
Here, the exponent -1 on $z$ is only for convenience. The operator product expansions for them are given by
\begin{align}
 &  \tilde {J}^{(+)}(z)\tilde {J}^{(-)}(z') = \frac{1}{(z-z')^2} - \frac{\omega }{z(z-z')}~,
 \label{eq:J+J-} \\
 &  \tilde {J}^{(-)}(z)\tilde {J}^{(+)}(z') = \frac{1}{(z-z')^2} + \frac{\omega }{z'(z-z')}~,
 \label{eq:J-J+} \\
 &   \tilde {I}^{(+)}(z)\tilde {I}^{(-)}(z') = \frac{1}{(z-z')^2} + \frac{\omega }{z'(z-z')}~,
 \label{eq:I+I-} \\
 &  \tilde {I}^{(-)}(z)\tilde {I}^{(+)}(z') = \frac{1}{(z-z')^2} - \frac{\omega }{z(z-z')}~,
 \label{eq:I-I+}
\end{align} 
Here, we have used the following contractions:
\begin{align}
 & \left\langle~{\alpha _m}^{(\pm)}{\alpha _n}^{(\mp)}  ~\right\rangle =
  \left\{\begin{array}{rl}
        (m \mp \omega )\delta _{m+n,0},& \quad (m\mp\omega >0) \\
        0,& \quad (m \mp \omega <0)
         \end{array}\right.  
 \label{eq:alpha_alpha} \\
 & \left\langle~{\beta _m}^{(\pm)}{\beta _n}^{(\mp)}  ~\right\rangle =
  \left\{\begin{array}{rl}
        (m \pm \omega )\delta _{m+n,0},& \quad (m\pm\omega > 0) \\
        0,& \quad (m \pm \omega <0)
         \end{array}\right.
 \label{eq:beta_brta} 
\end{align}
For the fermionic fields we confine ourselves to the NS sector,
\begin{align}
 &  \psi ^{(\pm)}(z) = \half z^{\mp\omega } \sum _r z^{-r-1/2} {b_r}^{(\pm)} = z^{\mp \omega }\tilde {\psi }^{(\pm)}(z)~,
 \label{eq:deftildePsi} \\
 &  \tilde {\psi }^{(\pm)}(z)\tilde {\psi }^{(\mp)}(z') = \frac{1/4}{z-z'}~,
 \label{eq:PsiPsi2}
\end{align}
The exponent -1/2 on $z$ in Eq.(\ref{eq:deftildePsi}) is only for convenience. \\
\indent Define the super current operator for the relevant part by
\begin{align}
 &  G(z) = 2 \sum^2_{\mu =1} \psi _{\mu }(z) I^{\mu }(z) = 2\big[ \psi ^{+}(z) I^{(-)}(z) + \psi ^{-}(z) I^{(+)}(z) \big] 
   \nonumber \\
 &      = 2\big[ \tilde {\psi }^{+}(z) \tilde {I}^{(-)}(z) + \tilde {\psi }^{-}(z) \tilde {I}^{(+)}(z) \big]~.
 \label{eq:defG}
\end{align}
Then we calculate the operator product $G(z)G(z')$ to yield the conformal operator $T_{-}(z)$, i.e.,
\begin{align}
 &  G(z)G(z') = \frac{2}{(z-z')^3} + \frac{\omega }{zz'(z-z')} + \frac{2T_{-}(z')}{z-z'}~,
 \label{eq:GG'}
\end{align}
where
\begin{align}
 &  T_{-}(z) = \half : I_{\mu }I^{\mu } : + 2: \partial \psi _{\mu } \psi ^{\mu } : = T_{-}^B(z) + T_{-}^F(z)~.
 \label{eq:TT'}
\end{align}
In the same way, for the bosonic part $T_{-}^B=:\tilde {I}^{(+)}\tilde {I}^{(-)}:$, we get
\begin{align}
 &  T_{-}^B(z)T_{-}^B(z') = \frac{1}{(z-z')^4} + \frac{\omega - \omega ^2}{zz'(z-z')^2} + \frac{2T_{-}^B(z')}{(z-z')^2} + \frac{\partial 'T_{-}^B(z')}{z-z'}~,
 \label{TT'}
\end{align}
and, for the fermionic part $T_{-}^F=2:\partial \psi \cdot \psi :$ with
\begin{align}
 &  \half T_{-}^F = : \partial \psi ^{(+)} \psi ^{(-)} + \partial \psi ^{(-)} \psi ^{(+)} :~
 \nonumber \\
 &                = : -\frac{2\omega }{z}\tilde {\psi }^{(+)} \tilde {\psi }^{(-)} +
 \partial \tilde {\psi }^{(+)} \tilde {\psi }^{(-)} + \partial \tilde {\psi }^{(-)} \tilde {\psi }^{(+)} :~,
 \label{eq:Ff}
\end{align}
we have
\begin{align}
   T_{-}^F(z)T_{-}^F(z') = \frac{\half}{(z-z')^4} + \frac{\omega ^2}{zz'(z-z')^2} + \frac{2T_{-}^F(z')}{(z-z')^2}
 + \frac{\partial 'T_{-}^F(z')}{z-z'}~.
 \label{eq:TFTF'}
\end{align}
Totally, it follows that
\begin{align}
   T_-(z)T_-(z') = \frac{3/2}{(z-z')^4} + \frac{\omega }{zz'(z-z')^2} + \frac{2T_{-}(z')}{(z-z')^2}
 + \frac{\partial 'T_{-}(z')}{z-z'}~.
 \label{eq:TT'2}
\end{align}
The algebra is closed by
\begin{align}
   T_-(z)G(z') = \frac{3/2}{(z-z')^2}G(z') + \frac{1}{z-z'}\partial 'G(z') ~.
 \label{eq:T-G'}
\end{align}
Finally, the algebra is supplemented by the + mode operator $T_{+}(z)=:\tilde {J}^{(+)}(z)\tilde {J}^{(-)}(z):$,
\begin{align}
   T_+(z)T_+(z') = \frac{1}{(z-z')^4} + \frac{\omega -\omega ^2}{zz'(z-z')^2} + \frac{2T_+(z')}{(z-z')^2}
 + \frac{\partial 'T_+(z')}{z-z'}~.
 \label{eq:T+T+'}
\end{align}
\\
\indent We have so far considered only the $(1, 2)$ plane, where the constant magnetic field is placed. The other ($d-2$)-components of space-time are free, and well known. Collecting all components of space-time, we get
\begin{align}
 &  T_+(z)T_+(z') = \frac{d_+/2}{(z-z')^4} + \frac{\omega -\omega ^2}{zz'(z-z')^2} + \frac{2T_+(z')}{(z-z')^2}
 + \frac{\partial 'T_+(z')}{z-z'}~,
 \label{eq:T+T+'2} \\
 &  T_-(z)T_-(z') = \frac{3d_+/4}{(z-z')^4} + \frac{\omega }{zz'(z-z')^2} + \frac{2T_-(z')}{(z-z')^2}
 + \frac{\partial 'T_-(z')}{z-z'}~,
 \label{eq:T-T-'2} \\
 &  T_-(z)G(z') = \frac{3/2}{(z-z')^2}G(z') + \frac{1}{z-z'}\partial 'G(z') ~.
 \label{eq:T-G'2} \\
 &  G(z)G(z') = \frac{d_-}{(z-z')^3} + \frac{\omega }{zz'(z-z')} + \frac{2T_-(z')}{z-z'}~.
 \label{eq:GG'2} 
\end{align}
These are equivalent to the super Virasoro algebra
\begin{align}
 & \big[\, L_{m}^{+}\,,~L_{n}^{+}\, \big] = (m-n)L_{m+n}^+ + \delta _{m+n,0}A_m^+~,
\label{eq:L+L+CCR} \\
 & \big[\, L_{m}^{-}\,,~L_{n}^{-}\, \big] = (m-n)L_{m+n}^- + \delta _{m+n,0}A_m^-~,
\label{eq:L-L-CCR} \\
 & \big[\, L_{m}^{-}\,,~G_r\, \big] = \big(\frac{m}{2}-r\big)G_{m+r}~,
\label{eq:L-GrCCR} \\
 & \big\{\, G_r\,,~G_s\, \big\} = 2L_{r+s}^- + \delta _{r+s,0}B_r^-~,
\label{eq:GrGrCCR} 
\end{align}
where the anomaly terms are given by
\begin{align}
 &  A_m^+ = \frac{d_+}{12}m(m^2-1) + m(\omega -\omega ^2)~,
 \label{eq:Am+} \\
 &  A_m^- = \frac{d_-}{8}m(m^2-1) + m\omega ~,
 \label{eq:Am-} \\   
 &  B_r^- = \frac{d_-}{2}\big(r^2-\frac{1}{4}\big) + \omega ~,
 \label{eq:Br-} \\
 &  B = q+\omega , \quad q\in Z, \quad 0<\omega <1~.
 \label{eq:defB_q}
\end{align}
%
\section{Anomaly in the Ramond sector} \label{sec:4}
As for the Ramond sector, we should be careful for the 0-mode. The mode expansions of fermionic fields are given by
\begin{align}
   {\psi _R}^{(\pm)}(z) = \half z^{-\omega } \sum_n z^{-n}d_n^{(\pm)} = z^{-\omega } \tilde {\psi }^{(\pm)}(z)~.
 \label{eq:Psipm}
\end{align}
where $n$ runs over the integral-numbers. The mode operators obey the commutation relation, after the Dirac quantization, 
\begin{align}
  \big\{\, d_m^{(+)}\,,~d_n^{(-)}\, \big\} = \delta _{m+n,0}~.
\label{eq:d-CCR2}
\end{align}
Usually the 0-mode $d_0^{\mu }$ is regarded as the Dirac  $\gamma $-matrix. However, in the presence of the magnetic field, 
it is not the case. The reason is as follows: Note that the super Virasoro operator $F_0$ contains factors, 
$\beta _0^{(+)}d_0^{(-)}+\beta _0^{(-)}d_0^{(+)}$. Since $\beta _0^{(-)}$ is the creation operator, 
the second term contradicts with the Virasoro condition $F_0\ket{\mbox{ground state}}=0$, if $d_0^{(\pm)}$ is regarded as the Dirac  $\gamma $ matrix. 
In the sector of the presence of magnetic fields, therefore, $d_0^{(+)}$ should be regarded as the annihilation operator, 
whereas other components $d_0^{\mu }$ without magnetic fields behave as $\gamma $ matrices. \\
\indent In this reason we consider that $d_m^{(+)}$ is annihilation operator for $m\geq 0$, 
and creation operator for $m<0$, while $d_m^{(-)}$ is annihilation operator for $m>0$, and creation operator for $m\leq 0$. 
The contractions are, therefore, defined as
\begin{align}
 & \left\langle~d_m^{(+)}d_n^{(-)} ~\right\rangle =
  \left\{\begin{array}{rl}
        \delta _{m+n,0},& \quad (m\geq 0) \\
        0,& \quad (m<0)
         \end{array}\right.  
 \label{eq:d+_d-} \\
 & \left\langle~d_m^{(-)}d_n^{(+)} ~\right\rangle =
  \left\{\begin{array}{rl}
        \delta _{m+n,0},& \quad (m>0) \\
        0,& \quad (m\leq 0)
         \end{array}\right.
 \label{eq:d-_d+} 
\end{align}
\\ 
The operator product expansions for fermionic fields are, then, given by
\begin{align}
 & \tilde {\psi }_R^{(+)}(z)\tilde {\psi }_R^{(-)}(z') = \frac{z}{4(z-z')}~,
 \label{eq:Psi+Psi-2} \\
 & \tilde {\psi }_R^{(-)}(z)\tilde {\psi }_R^{(+)}(z') = \frac{z'}{4(z-z')}~.
 \label{eq:Psi-Psi+2}
\end{align}
For the super operator, $F(z)=2[\tilde {\psi }_R^{(+)}(z)\tilde {I}^{(-)}(z)+\tilde {\psi }_R^{(-)}(z)\tilde {I}^{(+)}(z)]$, 
we have 
\begin{align}
  \mbox{anomaly terms of } F(z)F(z')=\frac{z+z'}{(z-z')^3}~.
 \nonumber
\end{align}
From the formula
\begin{align}
  \big\{\, F_m\,,~F_n\, \big\} = \oint dz dz'~z^{m}{z'}^{n} F(z)F(z')~,
\label{eq:defFmFn}
\end{align}
it follows that
\begin{align}
 & \big\{\, F_m\,,~F_n\, \big\} = 2L_{m+n}^- + \delta _{m+n,0}B_m^-~,
\label{eq:FmFn2} \\
 & B_m^-(\mbox{Ramond}) = \frac{d_-}{2}m^2~.
\end{align}
For the fermionic part $T_-^F=2:\partial \psi _R\cdot \psi _R:$ with
\begin{align}
 & \half T_-^F = :\partial {\psi _R}^{(+)}{\psi _R}^{(-)} + \partial {\psi _R}^{(-)}{\psi _R}^{(+)} : 
 \nonumber \\
 & = :-\frac{2\omega }{z} {\tilde {\psi }_R}^{(+)}{\tilde {\psi }_R}^{(-)} + \partial {\tilde {\psi }_R}^{(+)}{\tilde {\psi }_R}^{(-)} 
  + \partial {\tilde {\psi }_R}^{(-)}{\tilde {\psi }_R}^{(+)} :~.
 \label{eq:fermTF}
\end{align}
we have
\begin{align}
  \mbox{Anomaly parts of }T_-^F(z)T_-^F(z') = \frac{1}{(z-z')^4}\frac{z^2+z'^2}{4} + \frac{\omega ^2 - \omega }{(z-z')^2}~.
 \label{eq:anomalyTFTF'}
\end{align}
\indent For the bosonic part $T_-^B=:\tilde {I}^{(+)}\tilde {I}^{(-)}:$, we already had the product $T_-^B(z)T_-^B(z')$ before as
\begin{align}
  \mbox{Anomaly parts of } T_-^B(z)T_-^B(z') = \frac{zz'}{(z-z')^4} + \frac{\omega -\omega ^2}{(z-z')^2}~.
 \label{eq:anomalyT-T-}
\end{align}  
Here the equation has been multiplied by the factor $zz'$, in order to make it the same power as the fermionic one. 
Then the total sum of the anomaly is given by
\begin{align}
 & \mbox{Anomaly of } \big[T_-^B(z) + T_-^F(z)\big]\big[T_-^B(z') + T_-^F(z')\big] 
 \label{eq:anomalyT+T} \\ 
 &   = \frac{1}{(z-z')^4}\big( zz' + \frac{z^2 + z'^2}{4} \big)~.
 \nonumber
\end{align}  
where $\omega $ terms are cancelled out from Eqs.(\ref{eq:anomalyT-T-}) and (\ref{eq:anomalyT-T-}). 
This gives the anomaly term without the cyclotron frequency
\begin{align}
   A_m^-(\mbox{Ramond}) = \frac{d_-}{8}m^3~,
 \label{eq:Am2}
\end{align}
together with
\begin{align}
   B_m^-(\mbox{Ramond}) = \frac{d_-}{2}m^2~.
 \label{eq:Bm2}
\end{align}
%
\section{Spectrum-generating algebra} \label{sec:5}
Our SGA is characterized by the cyclotron frequency $\omega $. We summarize it for the right-moving NS sector:
\begin{align}
 & \big[\, A_m^{(+)}\,,~A_n^{(-)}\, \big] = (m+\omega )\delta _{m+n,0}~, \quad
  \big\{\, B_r^{(+)}\,,~B_s^{(-)}\, \big\} = \delta _{r+s,0}~,  \quad
  \big[\, A_m^i\,,~A_n^j\, \big] = 0~,
 \nonumber \\
 & \big[\, A_m^{(\pm)}\,,~A_n^+\, \big] = (m\pm \omega )A_{m+n}^{(\pm)}~, \quad
  \big[\, B_r^{(\pm)}\,,~A_n^+\, \big] = \big( \frac{n}{2} + r \pm \omega \big)B_{r+n}^{(\pm)}~, 
 \label{eq:SGA1} \\
 & \big[\, A_m^{(\pm)}\,,~B_r^+\, \big] = (m\pm \omega )B_{m+r}^{(\pm)}~, \quad
  \big\{\, B_r^{(\pm)}\,,~B_s^+\, \big\} = A_{r+s}^{(\pm)}~, 
 \nonumber
\end{align}
\begin{align}
 & \big[\, A_m^{+}\,,~A_n^{+}\, \big] = (m-n )A_{m+n}^{+} + m^3\delta _{m+n,0}~,
 \nonumber \\
 & \big[\, A_m^+\,,~B_r^+\, \big] = \big(\frac{m}{2} - r \big)B_{r+s}^+~,
 \label{eq:SGA2} \\
 & \big\{\, B_r^+\,,~B_s^+\, \big\} = 2A_{r+s}^{+} + 4r^2\delta _{m+n,0}~.
 \nonumber
\end{align}
Any operator in Eqs.(\ref{eq:SGA1}) and (\ref{eq:SGA2}) is commutable with the super Virasoro operator $G_r$.
Each definition of the operators in Eq.(\ref{eq:SGA1}) is as follows:
\begin{align}
 & A_m^{(\pm)} = \frac{1}{2\pi i} \oint dz A_m^{(\pm)}(z)~,
 \label{eq:defAmpm} \\
 & A_m^{(\pm)}(z) = \big[ I^{(\pm)} - (m\pm \omega )\psi ^{(\pm)}\psi _-\big]V^{m\pm \omega }~, \nonumber \\
 & V = :\exp{[iX_-(z)}]:~, \nonumber \\
 & X_-(z) = x_- - i p_- \ln z + i\sum_{n=0} \frac{\alpha _n^-}{n}z^{-n}~, \quad p_-=1~., \nonumber
\end{align}
\begin{align}
 & B_r^{(\pm)} = \frac{1}{2\pi i}\oint dz B_r^{(\pm)}(z)~,
 \label{eq:defBrpm} \\
 & B_r^{(\pm)}(z) = \big[\psi ^{(\pm)}\big( 1+\half \psi_-\partial \psi _-J_-^{-2} \big)J_-^{1/2} 
  - \psi _-I^{(\pm)}J_-^{-1/2} \big] V^{r\pm \omega }~. \nonumber
\end{align}
In Eqs.(\ref{eq:defAmpm})-(\ref{eq:defBrpm}) and Eqs.(\ref{eq:defAm+})-(\ref{eq:defBr+}) below, for brevity, 
fermionic fields are normalized in such a way that the contraction is given by 
$\left\langle~\psi ^{\mu }(z)\psi ^{\nu }(z') ~\right\rangle = (z-z')^{-1}\eta ^{\mu \nu }$. 
Here $\psi _-$, $X_-$ are light-cone variables defined by $X_{\pm}=\kappa ^{\pm 1}(X^0 \pm X^{d-1})/\sqrt{2}$, 
with a real parameter $\kappa $. The superscripts $(\pm)$ of $X^{(\pm)}=(X^{1} \pm i X^{2})/\sqrt{2}$ 
should be distinguished from the light-cone subscripts $\pm$. 
Note that the vertex operator $V^{m\pm \omega }=:\exp[i(m\pm\omega )X_-(z)]:$ behaves like $\sim z^{m\pm\omega }$ at $z=0$. 
The new definition for $A_m^{(\pm)}$ and $B_r^{(\pm)}$ reduce to the original ones proposed by Brower and Friedmann[5], 
if the  cyclotron frequency $\omega $ is set to be zero. \\
\indent The sub-algebra (\ref{eq:defBrpm}) is the same as that in Ref.\cite{ref:Brower_F}. 
It is composed only of the light-cone variables. They are free operators and this algebra is well-known. The light-cone operators are defined by
\begin{align}
 & A_m^{+} = \frac{1}{2\pi i} \oint dz A_m^{+}(z)~,
 \label{eq:defAm+} \\
 & A_m^{+}(z) = \big[ (J_- - n\psi _{+}\psi _-) - \half n(\partial J_-J_{-}^{-1} -n\psi _{-} \partial \psi _{-}J^{-1} \big]V^{n}~, 
 \nonumber
\end{align}
\begin{align}
 & B_r^{+} = \frac{1}{2\pi i}\oint dz B_r^{+}(z)~,
 \label{eq:defBr+} \\
 & B_r^{+}(z) = \big[\psi _{+}\big( 1+\half \psi_-\partial \psi _{-}J_{-}^{-2} \big)J_{-}^{1/2} 
  - \psi _{-}J_{+}J_{-}^{-1/2} \big] V^{r} + \mbox{(irrelevant term)}~.
 \nonumber
\end{align}
The proof of our SGA (\ref{eq:SGA1}) is given by the same method as in Ref.\cite{ref:Kokado_KS}.
%
\section{Isomorphisms} \label{sec:6}
The algebra (\ref{eq:SGA2}) is similar to the super Virasoro algebra for transverse operators
\begin{align}
 & \big[\, L_m^{T}\,,~L_n^{T}\, \big] = (m-n)L_{m+n}^T + A^{T}(m)\delta _{m+n,0}~,
 \nonumber \\
&  \big[\, L_m^{T}\,,~G_r^{T}\, \big] = \big( \frac{m}{2}-r)G_{m+r}^{T}~,
\label{eq:LTGrCCR} \\
&  \big[\, G_r^{T}\,,~G_s^{T}\, \big] = 2L_{r+s}^{T} + B^{T}(r)\delta _{r+s,0}~,
 \nonumber 
\end{align}  
where
\begin{align}
 &  A^{T}(m) = \frac{d_{-}-2}{8}m(m^2-1) + 2ma_{-} + m\omega ~,
 \label{eq:ATm} \\
 &  B^{T}(r) = \frac{d_{-} - 2}{2}\big (r^2 - \frac{1}{4} \big ) + 2a_{-} + \omega ~,
 \label{eq:BTm} \\
 &  B = q+\omega , \quad q\in Z, \quad 0<\omega <1~.
 \label{eq:Beq2}
\end{align}
Here the superscript $T$ means that the operators are constructed from $L_m^{-}, G_r^{-}$, 
leaving oscillators with spacial components $\mu =1,2,\cdots, d-2$. The constant $a_{-}$ 
is included in $L_{m}^{T}$ as $-a_{-}\delta _{m,0}$. \\
\indent  The isomorphisms 
\begin{align}
  A_{m}^{+} \sim L_{m}^{T}, \quad B_{r}^{+} \sim G_r^{T}
 \label{eq:AmLTBrGT}
\end{align}
are completed, if there hold equations
\begin{align}
 &  A^{T}(m) = \frac{d_{-}-2}{8}m(m^2-1) + 2ma_{-} + m\omega = m^3~,
 \label{eq:ATm2} \\
 &  B^{T}(r) = \frac{d_{-} - 2}{2}\big (r^2 - \frac{1}{4} \big ) + 2a_{-} + \omega =4r^2~,
 \label{eq:BTm2} 
\end{align}
These two equations are consistent to give the solution,
\begin{align}
 &  d_{-} = 10~,
 \label{eq:d-} \\
 & a_{-} = \half (1-\omega )~.
 \label{eq:a-}
\end{align}
As for the Ramond sector, we have $d^{R}_{-} = 10$ and $a^{R}_{-}=0$. \\
\indent  The isomorphisms (\ref{eq:AmLTBrGT}) are also extended to other components interacting 
with the magnetic field. The algebra (\ref{eq:AmLTBrGT}) is similar to
\begin{align}
 & \big[\, \beta _m^{(+)}\,,~\beta _n^{(-)}\, \big] = (m+\omega )\delta _{m+n,0}~, \quad
  \big\{\, b_r^{(+)}\,,~b_s^{(-)}\, \big\} = \delta _{r+s,0}~,  \quad
  \big[\, \beta _m^i\,,~b_r^j\, \big] = 0~,
 \nonumber \\
&  \big[\, \beta _m^{(\pm)}\,,~L_n^T\, \big] = (m\pm \omega )\beta _{m+n}^{(\pm)}~, \quad
  \big[\, b_r^{(\pm)}\,,~L_n^T\, \big] = \big( \frac{n}{2} 
+ r \pm \omega \big)b_{r+n}^{(\pm)}~, 
 \label{eq:SGA1a} \\
& \big[\, \beta _m^{(\pm)}\,,~G_r^T\, \big] = (m\pm \omega )b_{m+r}^{(\pm)}~, \quad
  \big\{\, b_r^{(\pm)}\,,~G_s^T\, \big\} = \beta _{r+s}^{(\pm)}~, 
 \nonumber 
\end{align}
The isomorphisms are now completed by
\begin{align}
   A_{m}^{(\pm)} \sim \beta _m^{(\pm)}, \quad B_r^{(\pm)} \sim b_r^{(\pm)}~.
 \label{eq:AmbetaBrbr}
\end{align}
\indent  As for the + (left-moving) mode, it contains only the bosonic string. The algebra 
is obtained from the superstring by neglecting all fermionic parts. SGA is given by
\begin{align}
 & \big[\, A_m^{(+)}\,,~A_n^{(-)}\, \big] = (m - \omega )\delta _{m+n,0}~,
 \nonumber \\
 & \big[\, A_m^{(\pm)}\,,A_n^{+}\, \big] = (m\pm \omega )A_{m+n}^{(\pm)}~,  
 \label{eq:SGA2a} \\
 & \big[\, A_m^{+}\,,~A_n^{+}\, \big] = (m - n )A_{m+n}^{+} + 2m^3\delta _{m+n,0}~.
 \nonumber 
\end{align}
The isomorphisms, $A_{m}^{(\pm)}\sim \alpha _{m}^{(\pm)}$ and $A_{m}^{+}\sim L_{m}^{T}$, 
are completed if there holds the equation
\begin{align}
   A_{+\mbox{mode}}^{T}(m) = \frac{d_{+}-2}{12}(m^3-m) + 2ma_{+} + m(\omega -\omega ^2) = 2m^3~.
 \label{eq:A+mode}
\end{align}
From this we have
\begin{align}
 &  d_{+} = 26~,
 \label{eq:d+a+} \\
 &  a_{+} = 1- \frac{\omega - \omega ^2}{2}~.
 \nonumber 
\end{align} 
Any physical state should be satisfied by the BRST charge condition 
 $Q_{\mbox{BRST}}\ket{\mbox{phys.}}=0$, or equivalently by the super Virasoro conditions, 
$G_{r>0}\ket{\mbox{phys.}}=0$, $(L_{n\geq 0}^{-}-\delta _{n,0}a_{-})\ket{\mbox{phys.}}=0$, 
$(L_{n\geq 0}^{+}-\delta _{n,0}a_{+})\ket{\mbox{phys.}}=0$, for the NS sector, 
and $F_{n\geq 0}\ket{\mbox{phys.}}=0$, $L_{n\geq 0}^{-}\ket{\mbox{phys.}}=0$,
 $(L_{n\geq 0}^{+}-\delta _{n,0}a_{+})\ket{\mbox{phys.}}=0$ for the Ramond sector. 
It is well known that such physical states can be constructed from 
the SGA operators\cite{ref:Brower_F}.
%
\section{Concluding remarks}\label{sec:conclude}
We have proposed a new type of interaction of the electromagnetic field with the heterotic 
string. When the charged heterotic string is placed in the constant magnetic field $B$, 
we have shown that the system can be solved exactly, so as to be translated 
into the free type of heterotic string. However, this free type of closed string is not 
periodic at the boundary $\sigma =2\pi $, but yields the phase factors $\exp{(\pm2\pi i\omega )}$
, where $\omega $ is the cyclotron frequency. This causes the fact that the cyclotron 
frequency $\omega $ is included in all orders in mode expansions of the string and also in 
quantization conditions for mode operators. This differs from the completely free case. 
Therefore, our next task has been to calculate the superconformal algebra together with 
anomalies, and also to give the spectrum-generating algebra, which is necessary to 
construct physical states satisfying the Virasoro conditions. Finally, we point out that our 
system is equivalent to the completely free system when $B$ takes integral values. \\
\indent  Any gauge symmetry derived from the internal space $I=10, \cdots, 25$ is broken by the 
cyclotron frequency $\omega $. This comes from the constraint condition for the 
internal momentum
\begin{align}
   \sum_{I} p_{I}^2 = 1+\omega ^2+2(R^{-}-R^{+})~,
 \label{eq:constrainPi}
\end{align}
where $R^{\pm}$ are number operators. The internal momentum is expressed as
\begin{align}
   p_{I} = n_{I}R_{I}~, \quad (n_{I}\in Z)
 \label{eq:internalPi}
\end{align}
where $R_{I}$ is the $I$th-radius of the torus. Since $2(R^{-}-R^{+})$ takes integral values 
plus integral times of $2\omega $, we have
\begin{align}
   \sum_{I} P_{I}^2 = \sum_{I} n_{I}^2R_{I}^2 = N + 2N'\omega + \omega ^2~. \quad (N, N' \in  Z)
 \label{eq:TotalPi}
\end{align}
This can be regarded as the constraint for $R_{I}$. However, the Kac-Moody algebra is 
related only with $\sum_{I}p_{I}^2=$ integral value, so that any internal gauge symmetry 
is violated by $\omega $.  \\
\indent  When $\omega =0$, the external magnetic field $B$ takes an integral value. In this case, 
as already noted before, our system is equivalent to completely free system, and we have 
the well-known internal gauge symmetries.
\begin{acknowledgments}
\indent We thank T. Okamura for valuable discussions.
\end{acknowledgments}

\end{document}